\documentclass[a4paper]{article}

\usepackage{INTERSPEECH2019}
\usepackage{multirow}

\title{Unsupervised Domain Adaptation in Speech Recognition using Phonetic Features}
\name{Rupam Ojha, C Chandra Sekhar}
\address{Department of Computer Science and Engineering, IIT Madras}
\email{rupam@cse.iitm.ac.in, chandra@cse.iitm.ac.in}

\begin{document}

\maketitle
\begin{abstract}
  Automatic speech recognition is a difficult problem in pattern recognition because several sources of variability exist in the speech input like the channel variations, the input might be clean or noisy, the speakers may have different accent and variations in the gender, etc. As a result, domain adaptation is important in speech recognition where we train the model for a particular source domain and test it on a different target domain. In this paper, we propose a technique to perform unsupervised gender-based domain adaptation in speech recognition using phonetic features. The experiments are performed on the TIMIT dataset and there is a considerable decrease in the phoneme error rate using the proposed approach. 
\end{abstract}


\section{Introduction}
Automatic speech recognition is a difficult problem in pattern recognition because several sources of variability exist in the speech input. The sources of variability include the channel variations, the input might be clean or noisy, the speakers may have different accent and/or gender etc. The models trained for a particular language/gender might not perform well for another language/gender. As a result, domain adaptation is important in speech recognition where we train the model for a particular source domain and test it on a different target domain. 

One of the ways to deal with the differences which occur due to change in domains is to include the phonetic features in the speech recognition process. Traditional systems usually build models for subword units like phonemes or syllables. The phonemes may vary across languages but the phonetic features, which depend on the speech production mechanism of humans, remain the same across languages. Sinicalchi et. al\cite{siniscalchi2012experiments} captured this to build models which are portable across languages. The phonetic features also help in recognition of hyper-articulated speech or non-native speech \cite{metze2005articulatory}.
While expressing phonemes in acoustic domain like done in traditional speech recognition systems we can have various unexplained variations among the phonemes depending upon the context. The variations can be explained by expressing the phonemes in phonetic domain using the phonetic features.
Each of the phonemes has some critical articulators and non-critical articulators. The critical articulator remains the same, however non-critical articulators may change depending upon the context \cite{king2000detection}.
The importance of using phonetic features for building robust models in case of noisy speech was done by Kirchhoff \cite{kirchhoff1999robust} and Mitra et. al \cite{mitra2011articulatory}.
Phonetic features are also used for detecting different emotions and state of depression \cite{mitra2015effects}. 
These studies motivated us to use phonetic features for the purpose of domain adaptation. 
In this paper, we explore ways to include phonetic features in gender based domain adaptation to improve the speech recognition accuracy. The techniques proposed here can be extended to perform domain adaptation for different languages and clean/noisy datasets too. 
This is because phonetic features have already shown promising results in cross-language and noisy datasets \cite{siniscalchi2012experiments,kirchhoff1999robust,metze2005articulatory}.

There are several ways to define phonetic features. In King and Taylor \cite{king2000detection} a $13$ dimensional SPE classification system is used. The silence is also taken as one of the phonetic features. Each of the phonemes can be expressed in terms of one or more of the phonetic features. This motivated us to treat the detection of phonetic features as a multi-label classification problem. In this paper, we perform domain adaptation using phonetic features and further use them to perform phoneme recognition.

\textbf{Our contributions} : a) A theoretical proof for using the domain adversarial neural networks for the purpose of multilabel domain adaptation. b) Proposed a domain adaptation based method for performing unsupervised domain adaptation using the multilabel domain adversarial neural networks. The rest of the paper is organized as follows: Section 2 gives an overview of approaches to domain adaptation used in speech recognition. Section 3 describes the proposed approach in detail. Section 4 gives details of the experiments and results, followed by conclusion in section 5.

\section{Approaches to domain adaptation in speech recognition}
In speech recognition, the gender based domain adaptation is done with source domain as speech input from male speakers and target domain as female speakers or vice-versa. Approaches to explore the gender based domain adaptation for speech recognition include cycle-GAN \cite{zhu2017unpaired}, multi-discriminator cycle-GAN \cite{hosseini2018multi} and augmented cycle adversarial learning \cite{hosseini2018augmented}. 
An important contribution in the field of domain adaptation was Domain Adversarial Neural Networks (DANN) \cite{ganin2016domain} used for domain adaptation on sentiment analysis and image captioning datasets. For speech recognition, DANN was applied on clean speech as source domain and noisy speech as target domain of Aurora-4 corpus with excellent improvement \cite{sun2017unsupervised}.

The DANN was proposed to perform the domain adaptation for the single label classification. We will now describe DANN in brief.
In domain adaptation, given the input space $X$ and a set of labels $Y$, we train the models to perform the classification task. There are two types of distributions over $X \times Y$ called source domain and target domain. We have $n$ examples drawn i.i.d from source domain and $n'$ examples drawn i.i.d from target domain. In an unsupervised domain adaptation, the source domain examples have labels along with input while the target domain examples do not have labels. An easier case is semi-supervised domain adaptation, where a few of target domain examples also have labels. The goal of the classifier that is built is to perform the classification of target samples with low target risk. 
In several studies in domain adaptation the target error is bounded by the source error and a distance between the distributions. 

When both the source and target distributions are similar, then source risk is a good estimate of the target risk. The distance between the distributions can be measured in several ways. In DANN, $\mathcal{H}$-divergence based on work by Ben-David et al.\cite{ben2007analysis} is used for measuring distance between distributions.

Let us consider a feedforward neural network with a single hidden layer. Let the input space be formed by real numbers such that $X \subset \mathbb{R}^m$. The neural network contains input layer, hidden layer and output layer. The output layer predicts the class or label of the given input. The hidden layer learns a $D$-dimensional representation of the input. Let us denote the hidden layer by $H_f$ given by
\begin{equation}
    H_f(\mathbf{x}; \mathbf{W},\mathbf{b} ) = \text{sigmoid}(\mathbf{W} \mathbf{x} + \mathbf{b} )
\end{equation}
\noindent where $\mathbf{b}$ is the bias vector and
 $ \text{sigmoid}(\mathbf{x}) = \left[ \frac{1}{1+exp(-\mathbf{x}_i)} \right]_{i=1}^{|\mathbf{x}|} $. Instead of logistic sigmoid any activation function can be used.
The output layer takes as input the $D$-dimensional representation of hidden layer. Ley $O_y$ be the output layer function and is given by
\begin{equation}\label{lastlayer}
    O_y(H_f(\mathbf{x});\mathbf{V},\mathbf{c}) = \text{softmax}(\mathbf{V}H_f(\mathbf{x})+\mathbf{c})
\end{equation}

\noindent where $\mathbf{c}$ is the bias vector and $\text{softmax}(\mathbf{x}) = \left[ \frac{ \exp(x_i) }{\sum_{j=1}^{|\mathbf{x}|} \exp(x_j) } \right]_{i=1}^{|\mathbf{x}|} $ .
The output $O_y(.)$ denotes the probability of the particular class $y$. For an example $(\mathbf{x}_i,y_i)$ the loss function is the negative log-probability of the correct label.

\begin{equation}\label{lossfunction}
    L_y(O_y(H_f(\mathbf{x}_i)),y_i) = \log \frac{1}{O_y(H_f(\mathbf{x}))_{y_i}} = L_y^i(\mathbf{W,V,b,c})
\end{equation}
For the source domain we have the labels present so we can use the optimization problem as:
\begin{equation}\label{optimization}
    \underset{\mathbf{W,V,b,c}}{\min} \left[ \frac{1}{n} \sum_{i=1}^{n} L_y^i(\mathbf{W,V,b,c}) + \lambda R(\mathbf{W,b}) \right]
\end{equation}
\noindent Here the loss for the $i$-th example is the term $L_y^i$ and $R(.)$ is the regularization term.
The regularization term is based on the H-divergence \cite{ganin2016domain,ben2007analysis} which seeks to approximate the distance between source and target distributions. In order to define the regularization term a new layer called as \textit{domain classification layer} is used.
The output of the hidden layer $H_f$ can be viewed as an intermediate representation which is given as input to \textit{domain classification layer} represented by $O_d$.
This layer learns to differentiate the given input as coming from source or target domain and its function is given by
\begin{equation}
    O_d(H_f(\mathbf{x});\mathbf{u},z) = \text{sigmoid}(\mathbf{u}^T H_f(\mathbf{x}) + z )
\end{equation}
\noindent The loss of \textit{domain regressor} is given by
\begin{align}\nonumber
    L_d(O_d(H_f(\mathbf{x}_i)),d_i) = d_i \log \frac{1}{O_d(H_f(\mathbf{x}_i))} + \\
    (1-d_i) \log \frac{1}{1-O_d(H_f(\mathbf{x}_i))}
\end{align}
\noindent where $d_i$ denotes the domain label for the $i$-th example, which denotes that the example came from source domain($d_i=0$) or target domain($d_i=1$).

In an unsupervised domain adaptation, the labels($y_i \in Y$) of the samples in source  domain ($d_i=0$) are available during the training. However, the labels of the samples in the target domain ($d_i=1$) are not available during training and should be predicted at test time. The optimization problem in (\ref{optimization}) is modified by adding the following term for regularization:
\begin{align}\nonumber
&    R = \\
 &    \underset{\mathbf{u},z}{max} \left[ -\frac{1}{n}\sum_{i=1}^{n} L_d^i(\mathbf{W,b,u},z) -\frac{1}{n'}\sum_{i=n+1}^{N} L_d^i(\mathbf{W,b,u},z) \right]
\end{align}

The complete objective function that implements a trade-off between the minimization of divergence (i.e. the distance between the source and target distributions) and the risk for training model for source distribution is given by

\begin{align}\nonumber\label{fullobjective}
&  E(\mathbf{W,b,V,c,u},z)   \\
    =& \frac{1}{n} \sum_{i=1}^{n} L_y^i - 
\lambda (\frac{1}{n}\sum_{i=1}^{n} L_d^i + \frac{1}{n'}\sum_{i=n+1}^{N} L_d^i )
\end{align}
The optimized parameters $\mathbf{\hat{W},\hat{b},\hat{V},\hat{c},\hat{u}},\hat{z}$ constitutes a saddle point, i.e.,
\begin{align*}
    (\mathbf{\hat{W},\hat{b},\hat{V},\hat{c}}) =&  \underset{\mathbf{W,b,V,c}}{\arg \min}  E(\mathbf{W,b,V,c,\hat{u}},\hat{z}) \\
    (\mathbf{\hat{u}},\hat{z}) =&  \underset{\mathbf{u},z}{\arg \max}  E(\mathbf{\hat{W},\hat{b},\hat{V},\hat{c},u},z)
\end{align*}
The optimization problem is solved such that $E$ is minimized with respect to $\mathbf{W,b,V,c}$ and maximized with respect to $\mathbf{u},z$. This allows us to learn the hidden representation which is discriminatory with respect to the labels in the source domain but cannot distinguish between source and target domains. It enables the neural network to learn features which are discriminatory in nature irrespective of the domain.

The DANN was proposed for single label classification systems. A method to use DANN for multi-label classification is proposed in the next section. The multi-label DANN is used to perform domain adaptation of phonetic features and the output is used to perform gender based domain adaptation. The proposed model is described in detail in the upcoming sections.

\section{Proposed Method}
\begin{figure}[h]
  \centering
    \includegraphics[scale=0.15]{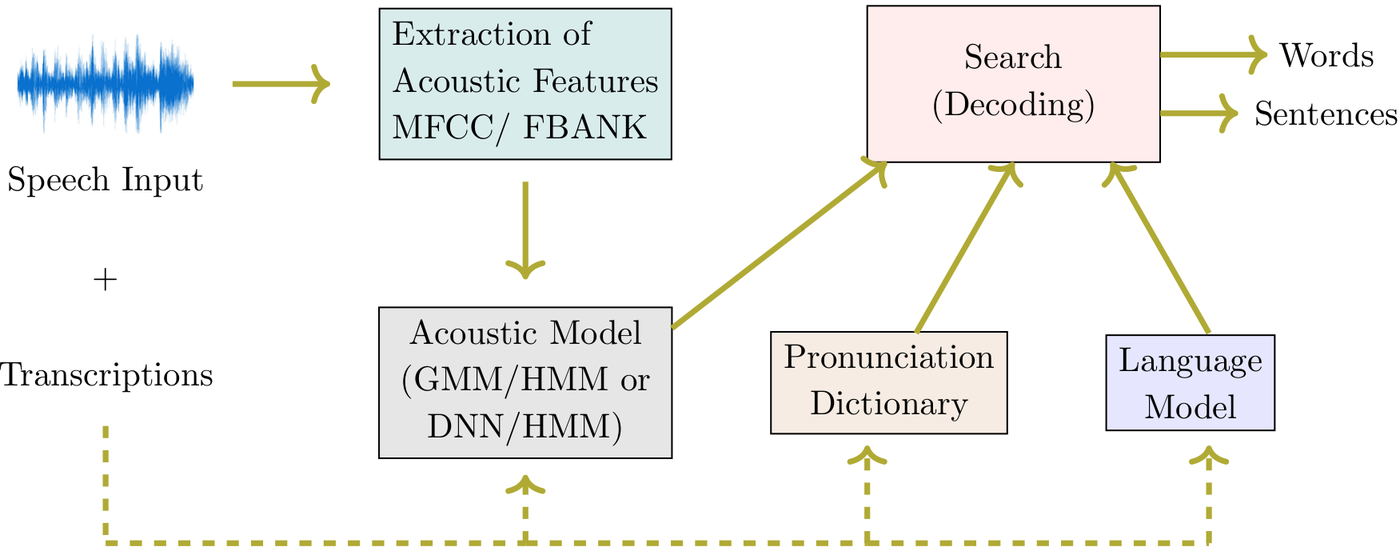}
    \caption{Conventional framework for ASR}
    \label{asr}
\end{figure}

The conventional framework of ASR is shown in figure \ref{asr}. The acoustic features are extracted from speech input and the acoustic models like Gaussian Mixture Model-Hidden Markov Model (GMM-HMM) or Deep Neural Network (DNN-HMM) are built. The DNN gives posterior probabilities corresponding to the phonemes. A search (decoding) is performed with the help of pronunciation dictionary and language models. In case of DNN-HMM systems, the posterior probabilities obtained by DNN are converted to likelihood before performing the decoding process. Using this framework, a model is trained on a source domain and tested on the target domain.

Domain adversarial neural networks (DANN) \cite{ganin2016domain} can be used to perform the domain adaptation task. A variant of DANN for noisy speech data was used in Sun et. al\cite{sun2017unsupervised}. We have used DANN to perform gender-based domain adaptation. The input to the DANN are the acoustic features and the desired output labels are the phonemes in a particular language. Since for a given frame only one of the phonemes can be associated with a given speech frame, we can apply the original DANN for the single label classification to obtain the phonemes. The DANN is trained in an unsupervised manner where the phoneme labels are used for training the model using the source domain data while unlabeled data is used from the target domain. Each of the source and target domain are given pseudo-labels as $0$ and $1$ respectively. The DANN is trained in an adversarial manner, so that it learns to discriminate between different phonemes irrespective of the domain. From a theoretical perspective, the task of DANN is to bring the source and target distributions closer to each other while also discriminating among the different classes. In an attempt to bring the source and target domains closer, the classification accuracy for the source domain decreases while that for the target domain increases as compared to the baseline models. In practical scenarios, we wish to increase the classification accuracy of the target domain, for which we do not have any labels during training time. However, we would also not like to compromise on the accuracy of the source domain in doing so. The proposed model achieves this by using the phonetic features to build a multi-label DANN.

\begin{figure}[ht]
    \centering
    \includegraphics[scale=0.20]{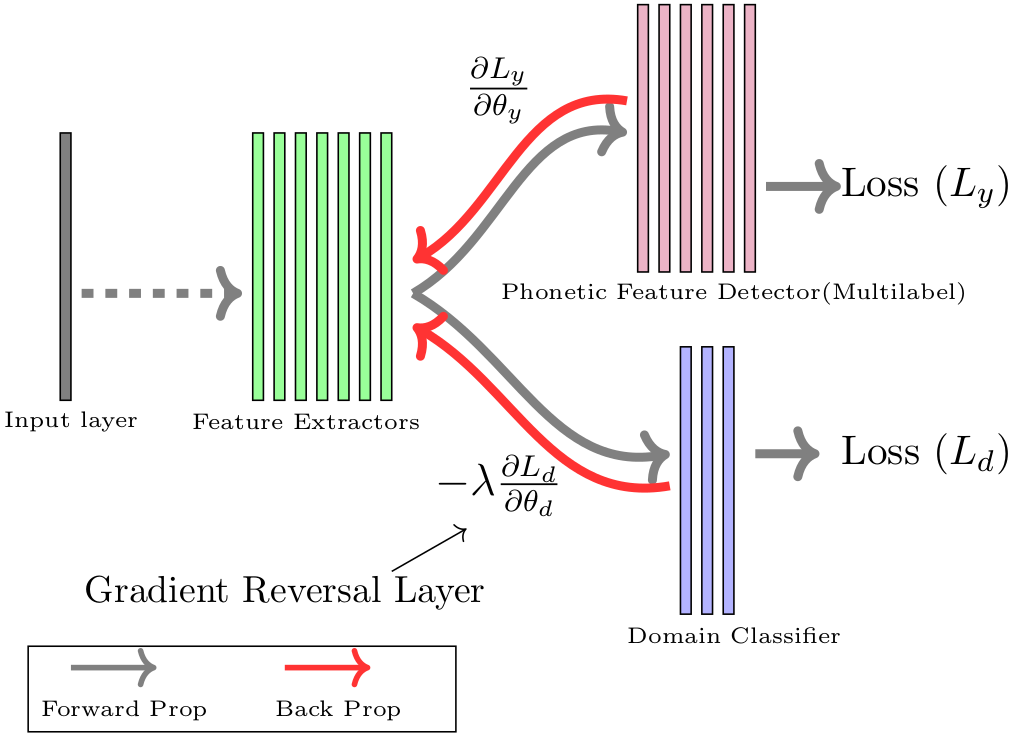}
    \caption{Mulit-label domain adversarial neural network}
    \label{dann}
\end{figure}
In multi-label classification, a given input may have more than one label present. Given an $m$-dimensional input, each input is associated with one or more class labels $L=\{0,1,\dots,L\}$. Each example can be associated with a binary vector of length $L$, with $0/1$ denoting the absence or presence of given label. Given $n$ such pairs of vectors $(\mathbf{x}_1,\mathbf{y}_1),(\mathbf{x}_2,\mathbf{y}_2),\dots,(\mathbf{x}_n,\mathbf{y}_n)$, we can use various approaches to train the model. The learning problem can be broken down into multiple independent binary classification problems where each example associated with a given label $y_i^j=1$ that will be considered as positive example for the $j$-th classifier and all others will be negative example.
While using neural networks in building models for multi-label classification, the global loss function is defined as
\begin{equation*}
    E = \sum_{i=1}^{n} E_i
\end{equation*}
\noindent where $E_i$ is the loss associated with the $i$-th example given by
\begin{equation}
    E_i = \sum_{j=1}^{L} (p_i^j-y_i^j)^2
\end{equation}
\noindent Here $p_i^j$ is the predicted output of the network for class $j$ with input $x_i$. The drawback of this approach is that it takes into account individual label discrimination i.e. whether the input belongs to a particular class and does not consider the correlations between the labels.
So, Zang et al. \cite{zhang2006multilabel} proposed a global loss function that is more appropriate for the multi-label data. It uses a pairwise error function (PWE) as given below
\begin{align}
    E =& \sum_{i=1}^{n} E_i \\
      =& \sum_{i=1}^n \frac{1}{|L_i||\overline{L}_i|} \sum_{(k,l) \in L_i \times \overline{L}_i} \exp (-(p_i^k-p_i^l))
\end{align}
It has been shown in Nam et al.\cite{nam2014large} that using the last layer as sigmoid layer and followed by cross-entropy loss performs better than PWE. 
For each input $\mathbf{x}_i$ the cross entropy loss is calculated as
\begin{equation}
    E_i = -\sum_{l} (y_i^l log p_i^l) + (1-y_i^l)log (1-p_i^l)
\end{equation}
Here, $y_i^l$ is the desired label and $p_i^l$ is the output for the label $l$.

For incorporating changes in the DANN architecture so that it can be used for multi-label datasets, the output layer (modified Equation (\ref{lastlayer})) $O_y: \mathbb{R}^D \rightarrow 2^L $ is learned as:
\begin{equation}
    O_y(H_f(\mathbf{x});\mathbf{V},\mathbf{c}) = \text{sigmoid}(\mathbf{V}H_f(\mathbf{x})+\mathbf{c})
\end{equation}
\noindent with $ \text{sigmoid}(\mathbf{x}) = \left[ \frac{1}{1+exp(-\mathbf{x}_i)} \right]_{i=1}^{|\mathbf{x}|} $. For each label $l$ the output is given by
$O_i^l$.
Also, the loss function in \ref{lossfunction} needs to be modified accordingly as
\begin{equation}\label{lossmultilabel}
    L_i(O_y(H_f(\mathbf{x}_i)),\mathbf{y}_i) = -\sum_l (y_i^l log O_i^l) + (1-y_i^l)log (1-O_i^l)
\end{equation}

Figure \ref{dann} shows the framework for multi-label DANN. It looks quite similar to the original DANN except the loss ($L_y$) of the phonetic feature detector is computed as specified in the Equation (\ref{lossmultilabel}) above. 

For a given phoneme, one or more phonetic features may be present. This enables us to formulate the detection of phonetic features as a multi-label classification problem. The phonetic features depend upon the speech production mechanism of the humans. So, it is easier to perform domain adaptation on phonetic features as compared to phonemes. The FBANK features are given as input to the multi-label DANN and the phonetic features labels are used as the desired output for calculating the loss in the phonetic feature detector model. The phonetic feature labels are obtained by the procedure described in King and Taylor \cite{king2000detection}. The multi-label DANN (Figure \ref{dann}) is trained in an adversarial manner with the labeled source domain examples and unlabeled target domain examples. The adversarial training of the  feature extractor enables it to learn the feature which can be used to discriminate phonetic feature classes while being invariant to the domain of the input.
The output of the DANN is considered as a probability score corresponding to each of the phonetic features. The procedure to obtain the phonetic feature labels and train the multi-label DANN is described in detail in the next section.

\begin{figure}[h]
  \centering
    \includegraphics[scale=0.19]{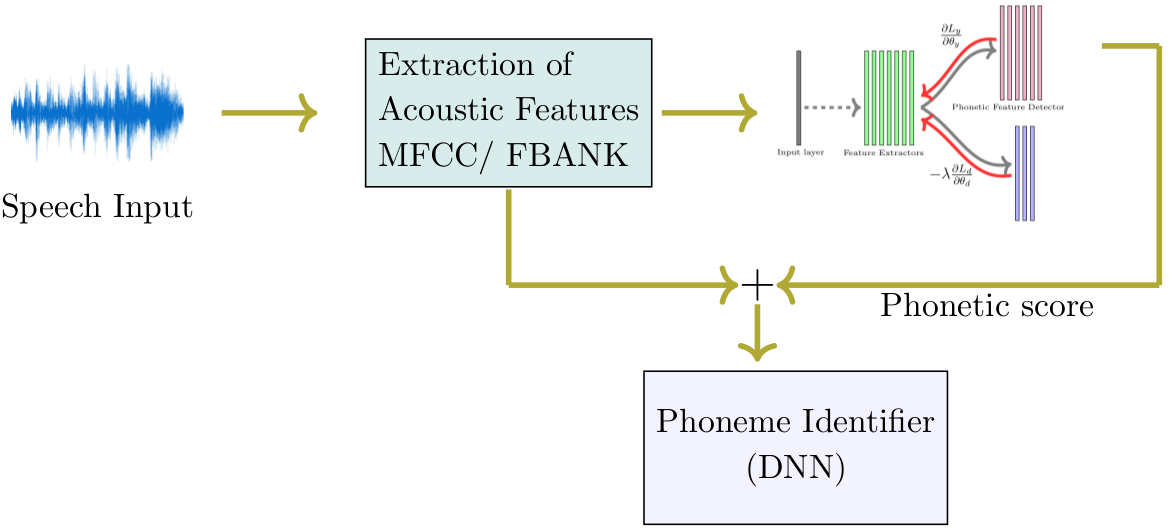}
    \caption{Proposed framework for domain adaptation. The phonetic score is obtained by the Multi-label DANN which is shown in Figure \ref{dann}.}
    \label{domain}
\end{figure}

In Figure \ref{domain}, we have shown the proposed framework for domain adaptation. This works well for both unsupervised as well as semi-supervised cases. We will describe the unsupervised case. The acoustic features are extracted for both source and target domains. The multi-label DANN is trained in an unsupervised manner for the source domain. The output of the phonetic feature classifier of the multi-label DANN is called the phonetic score. The phonetic score is calculated for the input in the source domain. This phonetic score is appended to the acoustic features and a DNN is trained for the identification of phonemes using the appended input. During test time, the phonetic score of the test data is taken from the already trained multi-label DANN and appended to the acoustic features. This appended feature vector is given as input to DNN to obtain the phonemes. Thus the entire framework is trained in an unsupervised manner without using the target domain labels. In order to use the target domain labels for the semi-supervised domain adaptation, we may train the multi-label DANN by providing input labels for the target domain and similarly train the phoneme identifier with the target domain labels.

\section{Experiments and Results}

The experiments were performed for male and female data of the TIMIT dataset \cite{garofolo1993darpa}. Only \textit{si} and \textit{sx} parts were used and \textit{sa} was removed from the TIMIT dataset. The number of train utterances were $3696$, out of which $206$ utterances were used as a validation set for determining the hyperparameters. The test set had $1344$ utterances. One of the genders was taken as the source domain and another as the target domain. In TIMIT dataset, the ratio of male data is around $70\%$ and female data is around $30\%$. The acoustic features extracted were using $23$ filter bank coefficients for interval of $25$ ms with a shift of $10$ ms. They were appended with delta and delta-delta features to obtain $69$ dimensional features. A context of $\pm 5$ frames was appended to obtain a segment of $11$ frames. 

In training the models for the source domain as male, only labels corresponding to the male utterances are used for training. The training is carried out in purely unsupervised manner and the labels of the target domain are used in test time to determine the performance.

The acoustic features are given as input to the modified DANN. The target labels for the DANN are the phonetic features. 
Since TIMIT is transcribed at phoneme level, the phonetic feature labels can easily be obtained using the SPE classification table as described in King and Taylor \cite{king2000detection}. The phonetic features in the SPE feature system are a set of $13$ phonetic features namely: vocalic, consonantal, high, back, low, anterior, coronal, round, tense, voice, continuant, nasal and strident. Silence is also taken as one of the features.
For datasets which do not have phoneme transcriptions, the phonetic feature labels can be obtained by training a triphone GMM-HMM model and doing viterbi decoding to obtain the phoneme alignments followed by mapping them to the correct phonetic feature targets. For a given source domain, we do not use the phonetic feature labels of target domain to train the multi-label DANN. The hyperparameters for efficient unsupervised training of the DANN are obtained by the \textit{reverse cross validation} approach \cite{ganin2016domain}. 

Once the multilabel DANN is trained, the output of the phonetic feature classifier is taken as the phonetic feature score. This score is appended with each of the $69$ dimensional FBANK acoustic features to obtain $69+14=83$ dimensional feature vector. The final DNN for recognition of phonemes is trained on a segment of $11$ frames of the appended features. The optimizer used for training the DNN is SGD and learning rate scheduling is performed. The results are reported in terms of phoneme error rate (PER) in Table \ref{my-label}.
\begin{table}[h]
\centering
\caption{Results (PER) of domain adaptation with training on either male or female data of TIMIT dataset.}
\label{my-label}
\begin{tabular}{|c|c|c|l|l|}
\hline
\multirow{2}{*}{\textbf{Model}} & \multicolumn{2}{c|}{\textbf{Train Male}}                                                                                      & \multicolumn{2}{l|}{\textbf{Train Female}}                                                                                      \\ \cline{2-5} 
                                & \textbf{\begin{tabular}[c]{@{}c@{}}Test\\ Male\end{tabular}} & \textbf{\begin{tabular}[c]{@{}c@{}}Test\\ Female\end{tabular}} & \textbf{\begin{tabular}[c]{@{}l@{}}Test \\ Male\end{tabular}} & \textbf{\begin{tabular}[c]{@{}l@{}}Test \\ Female\end{tabular}} \\ \hline
\textbf{GMM-HMM}               & 31.6                                                         & 44.7                                                           & 46.5                                                          & 34.7                                                            \\ \hline
\textbf{DNN-HMM}                    & 22.0                                                         & 37.8                                                           & 40.6                                                          & 26.9                                                            \\ \hline
\textbf{DANN\cite{ganin2016domain}}                   & 24.6                                                         & 35.0                                                           & 36.5                                                          & 29.8                                                            \\ \hline
\textbf{Proposed}               & \textbf{21.4}                                                & \textbf{35.0}                                                    & \textbf{35.9}                                                 & \textbf{24.0}                                                   \\ \hline
\end{tabular}
\end{table}

From Table \ref{my-label}, we can see that both GMM-HMM and DNN-HMM models fail to perform well when trained on data using a single gender. The DANN performs worse on the source domain but improves the recognition accuracy on the target domain. Our proposed method, though performs similar to DANN for the target domain, does significantly better in the source domain. This shows that we can improve the recognition accuracy in the target domain without compromising the performance for the source domain with the help of phonetic feature based domain adaptation.

\section{Conclusion}
Domain adaptation is important in ASR because several sources of variability exist while training the models for recognition of speech. In this paper, we proposed a model based on modified domain adversarial networks using the phonetic features. We showed that the proposed model seeks to improve the accuracy for the target domain without compromising the accuracy for the source domain. This approach is based on using the phonetic feature scores predicted by the modified DANN to build the models for phoneme recognition. As phonetic features have been shown to help building cross-language models and models for noisy speech, the proposed approach is expected to perform better for language based and clean/noisy speech domain adaptation experiments too.

\bibliographystyle{IEEEtran}
\bibliography{mybib}

\end{document}